\documentclass{aastex}
\usepackage{spr-astr-addons}
\usepackage{url}

\RequirePackage{color}

\newcommand{\emaila}{bltan@bao.ac.cn}

\begin{document}

\title{Multi-timescale Solar Cycles and the Possible Implications}
\shorttitle{Multi-solar Cycles} \shortauthors{Baolin Tan}

\author{Baolin Tan\altaffilmark{1}}
\affil{Key Laboratory of Solar Activity, National Astronomical
Observatories, Chinese Academy of Sciences.}

\email{\emaila}

\altaffiltext{1}{Datun Road A20, Chaoyang District, Beijing
100012, China, email: bltan@nao.cas.cn}

\begin{abstract}

Based on analysis of the annual averaged relative sunspot number
(ASN) during 1700 -- 2009, 3 kinds of solar cycles are confirmed:
the well-known 11-yr cycle (Schwabe cycle), 103-yr secular cycle
(numbered as G1, G2, G3, and G4, respectively since 1700); and
51.5-yr Cycle. From similarities, an extrapolation of forthcoming
solar cycles is made, and found that the solar cycle 24 will be a
relative long and weak Schwabe cycle, which may reach to its apex
around 2012-2014 in the vale between G3 and G4. Additionally, most
Schwabe cycles are asymmetric with rapidly rising-phases and
slowly decay-phases. The comparisons between ASN and the annual
flare numbers with different GOES classes (C-class, M-class,
X-class, and super-flare, here super-flare is defined as $\geq$
X10.0) and the annal averaged radio flux at frequency of 2.84 GHz
indicate that solar flares have a tendency: the more powerful of
the flare, the later it takes place after the onset of the Schwabe
cycle, and most powerful flares take place in the decay phase of
Schwabe cycle. Some discussions on the origin of solar cycles are
presented.

\end{abstract}
\keywords{Solar cycle --- Sun: extrapolation --- Sun: flares}


\section{Introduction}

It is well known that the Sun is an unique star which affect the
geo-space environment extremely. Any variations of solar
activities may greatly affect many sides of human living, from
satellite operations, radio-based communication, navigation
systems, electrical power grid, and oil tubes, etc. Therefore, it
is very important to understand when and how the solar activity
will take place in the near future. There are many people who had
made such endeavors (Hiremath, 2008; Hathaway, 2009; Strong,
Julia, \& Saba, 2009; etc). However, we have much of uncertainties
(Pesnell, 2008) to present the details of the forthcoming solar
cycles. The film \textbf{2012} reflects public misgivings to the
imminent impact of solar fierce eruptions around the year of 2012.

There are many methods to predict the solar cycle 24 and the
beyond. The Solar Cycle 24 Prediction Panel in October 2006 made a
comprehensive list of prediction methods including precursor,
spectral, climatology, recent climatology, neural network,
physics-based, etc (Pesnell, 2008). However, the investigation of
periodicity of solar activity is a groundwork for solar cycle
prediction. In previous works, many kinds of periodic modes are
found from the analysis of different solar proxies (e.g., yearly
averaged sunspot number, cosmogenic isotopes, historic records,
radiocarbon records in tree-rings, etc.), for example, the 11-yr
solar cycle, 53-yr period (Le \& Wang, 2003), 78-yr period (Wolf,
1862), 80 - 90-yr period (Gleissberg, 1971), the 65-130 yr
quasi-periodic secular cycle (Nagovitsyn, 1997), 100-yr periodic
cycle (Frick et al, 1997), 101-yr period cycle (Le \& Wang, 2003),
160 -- 270-yr double century cycle (Schove, 1979), 203-yr Suess
cycle (Suess, 1980). Otaola \& Zenteno (1983) proposed that long
term cycles within the range of 80 -- 100 and 170 -- 180 yr are
existed certainly.

Sunspot number is the most commonly predicted index of solar
activity. The number of solar flares, coronal mass ejections, and
the amount of energy released are well correlated with the sunspot
number. The present work mainly applied the annual averaged
relative sunspot number (ASN) during 1700 -- 2009 to study the
long-term variations, and found some meaningful insights of the
forthcoming solar solar activities. This paper is arranged as
following: section 2 presents the investigation of periodicity of
solar activity and their main features. Section 3 gives some
implications of solar cycles, such as the extrapolation of the
forthcoming Schwabe cycle 24, the distribution of the solar
flares, and the possible origin of solar cycles. Section 4 is the
conclusions and some discussions.

\section{The Periodicity of Solar Activity and its Main Features }

It has been realized for a long time that the solar activity is
connected tightly with the solar magnetic field. The energy
released in solar eruptive processes is coming from the magnetic
field. Hence the variations of solar magnetic field will be a
physically reasonable indicator for the solar activity. The
sunspot number presents a perceptible feature of the solar
magnetic field, which can be regarded as a commonly predicted
solar activity index. So, in this work, we adopt ASN during 1700
-- 2009 to investigate the periodicity of the solar long period
activities. The data set is downloaded from the internet:
http://sidc.oma.be/sunspot-data/.

The upper panel in Fig.1 presents the profile of ASN during 1700
-- 2009, and the lower panel is a Fourier power spectra of the
fast Fourier transformation (FFT) from ASN. Here presents 3
obvious spectral peaks which indicate that the strongest
periodicity is occurred at 11 years ($P_{0}$), the second
strongest periodicity is at 103 years ($P_{1}$), and the mildly
periodicity is at a period of 51.5 years ($P_{2}$).

\subsection{11-yr cycle ($P_{0}$)}

The most remarkable feature in Fig.1 is the solar cycles with
about 11-yr periods, it is the well-known Schwabe cycle. As ASN
reaches to the minimum 2.9 in 2008, and then increases slightly to
3.1 in the year of 2009. After coming into 2010, it becomes
stronger obviously than the past two years, there occurred several
GOES M-class flares in the first two months. So we may confirm
that the end of solar cycle 23 and the start of solar cycle 24 is
occurred around 2008. There are 28 Schwabe cycles during 1700 --
2009. They are numbered by an international regulation since 1755.
The solar cycle 1 started from the year of 1755, the solar cycle
23 ended around 2008. During 1700 -- 1755, there are 5 solar
Schwabe cycles. We may numbered them as $A$, $B$, $C$, $D$, and
$E$ respectively, which presented in the upper panel of Fig.1.

\begin{figure}
\begin{center}
 \includegraphics[width=8.0cm]{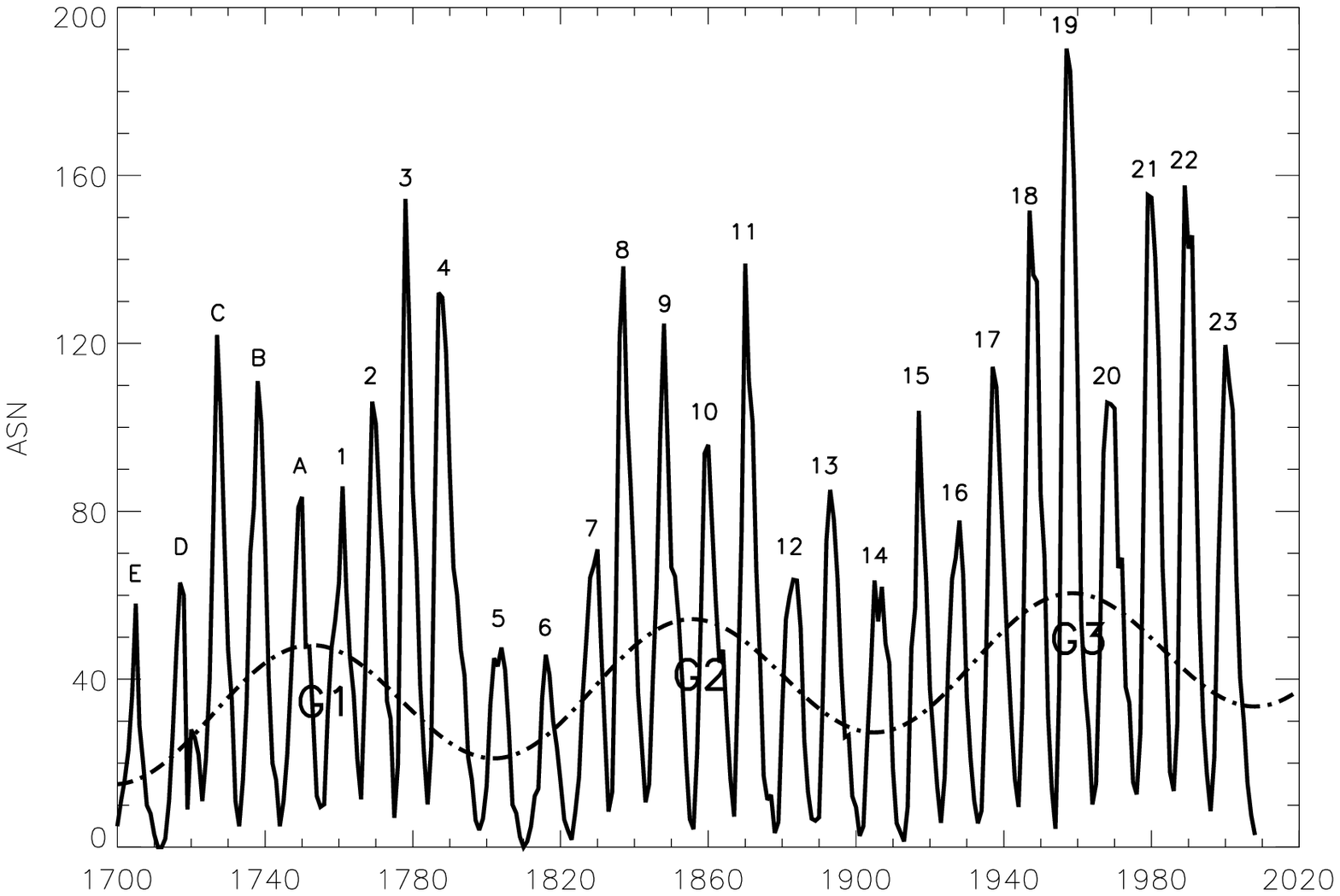}
 \includegraphics[width=8.0cm]{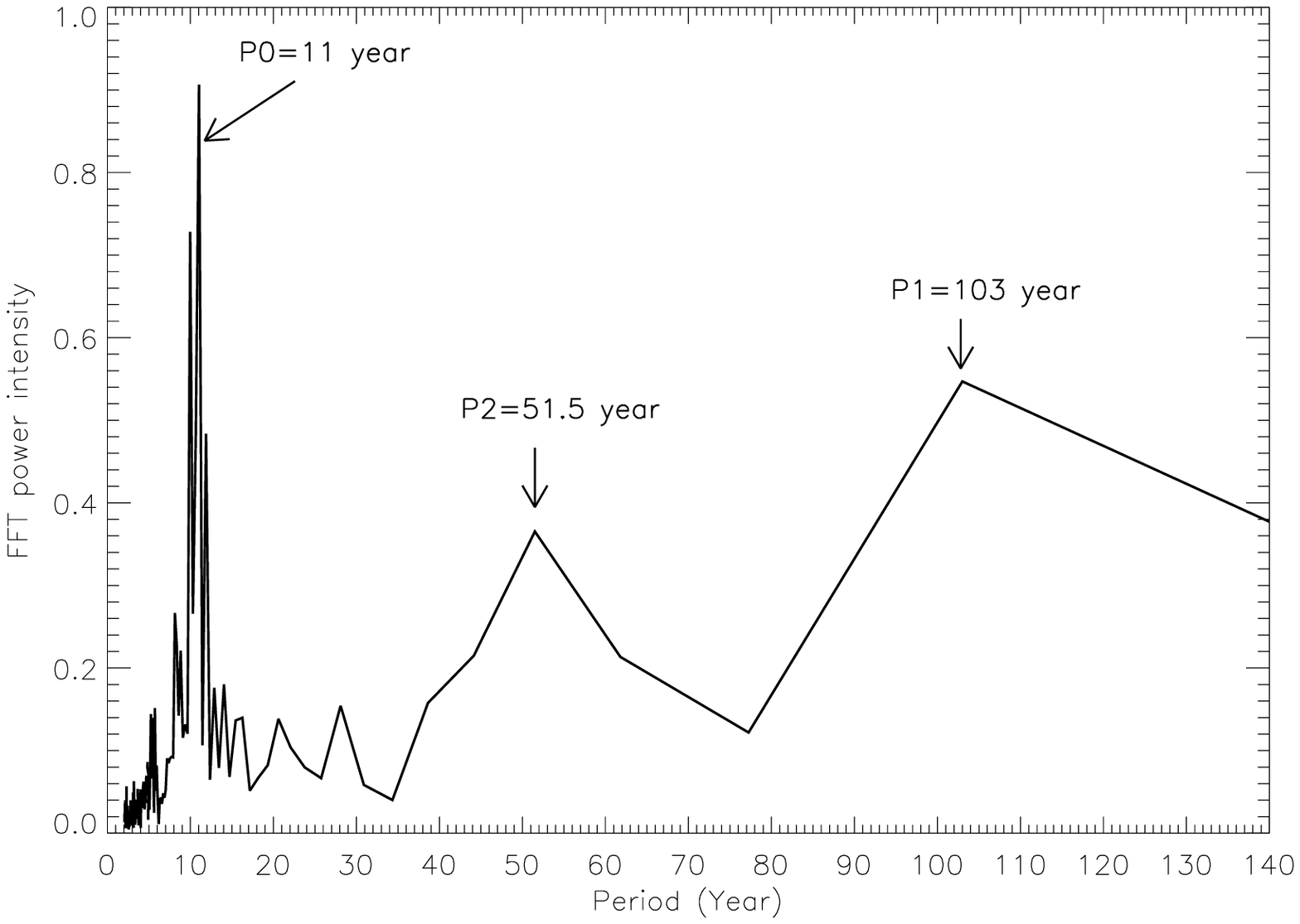}
  \caption{The upper panel is temporal profile of ASN during 1700 -- 2009 (solid curve).
  The thick dot-dashed curve is a fitted function which shows the secular cycle with period of about 103 years.
  The lower panel is the corresponding Fourier power spectra of the Fast Fourier Transformation.}
\end{center}
\end{figure}

In order to investigate the profile of each solar Schwabe cycle
more detail to get more useful insights, we may define several
parameters:

(a) rise-time ($t_{u}$), defined as from the starting minimum to
the maximum of a solar cycle in unit of year;

(b) decay-time ($t_{d}$), defined as from the maximum to the next
minimum of a solar cycle in unit of year;

(c) the profile of solar Schwabe cycles is always asymmetric,
i.e., taking less time to rise to the maximum than reaching to the
next minimum. We may define an asymmetric parameter as
($Asy=t_{u}/t_{d}$). When $Asy=1$, the profile is symmetric;
$Asy<1$ is left asymmetric, $Asy>1$ is right asymmetric;

(d) maximum ASN ($N_{max}$), which may represents the ASN
amplitude of each Schwabe cycle.

Table 1 lists all parameters of Schwabe cycles since 1700. The
last two lines in the Table present the mean value and the
standard deviation of above parameters.

\begin{table}
 \caption{List of characteristic parameters of solar Schwabe cycles during 1700 -- 2009. $t_{start}$, $t_{u}$,
 $t_{d}$, P, $Asy$, $R_{max}$ is the start time, rise-time, decay-time, period, asymmetric parameter,
 and the maximum ASN, respectively. Av and dev are the averaged value and the standard deviation of the related parameters, respectively.
 The star $*$ marks the bottom of the each secular cycle.}
\label{T-simple}
\begin{tabular}{lccccccc}\hline                   
     Cycle     & $t_{start}  $ & $t_{u}$  &  $t_{d}$    &    P     &    $Asy$     &  $N_{max}$   \\\hline
    $  E^{*}$  & $  1700     $ &    5     &   6         &    11    &    0.83      &     58       \\
    $  D   $   & $  1711     $ &    6     &   6         &    12    &    1.00      &     63       \\
    $  C   $   & $  1723     $ &    4     &   6         &    10    &    0.67      &     122      \\
    $  B   $   & $  1733     $ &    5     &   6         &    11    &    0.83      &     111      \\
    $  A   $   & $  1744     $ &    6     &   5         &    11    &    1.20      &     83.4     \\
    $  1   $   & $  1755     $ &    6     &   5         &    11    &    1.20      &     85.9     \\
    $  2   $   & $  1766     $ &    3     &   6         &     9    &    0.50      &     106.1    \\
    $  3   $   & $  1775     $ &    3     &   6         &     9    &    0.50      &     154.4    \\
    $  4   $   & $  1784     $ &    3     &   11        &    14    &    0.27      &     132.0    \\
    $  5^{*}$  & $  1798     $ &    6     &   6         &    12    &    1.00      &     47.5     \\
    $  6   $   & $  1810     $ &    6     &   7         &    13    &    0.86      &     45.8     \\
    $  7   $   & $  1823     $ &    7     &   3         &    10    &    2.33      &     70.9     \\
    $  8   $   & $  1833     $ &    4     &   6         &    10    &    0.67      &     138.3    \\
    $  9   $   & $  1843     $ &    5     &   8         &    13    &    0.63      &     124.7    \\
    $  10  $   & $  1856     $ &    4     &   7         &    11    &    0.57      &     95.8     \\
    $  11  $   & $  1867     $ &    3     &   8         &    11    &    0.38      &     139.0    \\
    $  12  $   & $  1878     $ &    5     &   6         &    11    &    0.83      &      63.7    \\
    $  13  $   & $  1889     $ &    4     &   8         &    12    &    0.50      &      85.1    \\
    $  14^{*}$ & $  1901     $ &    4     &   8         &    12    &    0.50      &      63.5    \\
    $  15  $   & $  1913     $ &    4     &   6         &    10    &    0.67      &     103.9    \\
    $  16  $   & $  1923     $ &    5     &   5         &    10    &    1.00      &     77.8     \\
    $  17  $   & $  1933     $ &    4     &   7         &    11    &    0.57      &     114.4    \\
    $  18  $   & $  1944     $ &    3     &   7         &    10    &    0.43      &     151.6    \\
    $  19  $   & $  1954     $ &    3     &   7         &    10    &    0.43      &     190.2    \\
    $  20  $   & $  1964     $ &    4     &   8         &    12    &    0.50      &     105.9    \\
    $  21  $   & $  1976     $ &    3     &   7         &    10    &    0.43      &     155.3    \\
    $  22  $   & $  1986     $ &    3     &   7         &    10    &    0.43      &     157.6    \\
    $  23  $   & $  1996     $ &    4     &   8         &    12    &    0.50      &     119.6    \\
    $  24^{*}$ & $  2008     $ &   (5)    &   (7)       &   (12)   &   (0.71)     &     (56.3)   \\
     Av        & $  --       $ &   4.4    &   6.6       &    11    &    0.72      &     105.9    \\
     dev       & $  --       $ &   1.2    &   1.4       &    1.3   &    0.40      &     37.8     \\\hline
\end{tabular}
\end{table}

Table 1 indicates that the rise-time of Schwabe cycles is in range
of 3 -- 7 years (the longest rise-time is 7 years in cycle 7),
with the averaged length of 4.4 years, and the standard deviation
of 1.2 years; the decay-time of Schwabe cycles is in range of 3
--11 years (the longest decay-time is 11 years in cycle 4), with
the averaged length of 6.6 years, and the standard deviation of
1.2 years; the period of Schwabe cycles is in range of 9 -- 14
years, with the averaged length of 11 years, and the standard
deviation of 1.3 years.

\subsection{103-yr secular cycle ($P_{1}$)}

The distribution of ASN amplitudes of Schwabe cycles shows clearly
the existence of 103-yr cycle, we call such long-term cycle as
secular cycle. Gleissberg firstly implied the existence of secular
solar cycle, so we also call it as Gleissberg cycle (Gleissberg,
1939). In fact, the secular cycle is consistent with the
appearance rate of the Grand Minima, such as the Sp$\ddot{o}$rer
Minimum around 1500, the Maunder Minimum around 1700 (1645 --
1715, Eddy, 1983), the Dalton Minimum around 1800 (1790 -- 1820).
From the distribution of ASN amplitudes of Schwabe cycles we may
also find that around 1900 (Schwabe cycle 14) seems also a Grand
Minimum. In other word, each of the Grand Minima is possibly
occurred in a vale between two secular cycles.

In order to show clearly the secular cycles, we make a fitted
empirical function with sinusoidal shape by using a method similar
to the square-least-method: at first we assume the sinusoidal
function in following formation:

\begin{equation}
R_{0i}=A+B\cdot sin(\frac{2\pi}{C}y_{i}+D\pi)+Ey_{i}
\end{equation}

Here, $y_{i}$ represents the time from 1700 (with unit of year).
Then let the following sum to become minimum:

\begin{equation}
Q(A,B,C,D,E)=\sum [R_{i}-R_{0i}]^{2}\rightarrow Min.
\end{equation}

In above equations, $R_{oi}$ and $R_{i}$ present the values of the
fitted empirical function and the observations of ASN,
respectively. As Equation (1) is a nonlinear function, we couldn't
obtain the true values of parameters A, B, C, D, and E from the
standard square-least-method. However, we may try to list a series
of [A, B, C, D, E] and calculate the values of $Q(A,B,C,D,E)$,
respectively. Then find out the minimum value of $Q(A,B,C,D,E)$
and the corresponding parameters of [A, B, C, D, E]. The thick
dot-dashed curve in the upper panel of Fig.1 is the fitted
empirical function, which can be expressed as:

\begin{equation}
R_{0}\approx 15[2.00+sin(\frac{2\pi}{103}y+\frac{3}{2}\pi)]+0.06y.
\end{equation}

From the empirical function we may find that the period of
long-term variation of ASN is 103 years, which is very close to
the secular cycle. We may numbered the secular cycles as $G1$
(includes the Schwabe cycle E -- A, and 1 -- 5), $G2$ (includes
the Schwabe cycle 6 -- 14), $G3$ (includes the Schwabe cycle 15 --
24), and $G4$ (after the Schwabe cycle 24) since 1700,
respectively marked in Fig. 1. At present the Sun is in a vale
between $G3$ and $G4$. The last term in the right hand of Equ.(3)
implies that secular cycles have a gradually enhancement, and the
Sun may have a tendency to become more and more active at the
timescale of several hundred years.

Many previous studies also presented the evidences of the secular
cycles (Nagovitsyn, 1997; Frick et al, 1997; Le \& Wang, 2003;
Bonev, Penev, \& Sello, 2004; Hiermath, 2006, etc.). The Grand
Minima (e.g., Sp$\ddot{o}$rer Minimum, Maunder Minimum, Dalton
Minimum, etc) implies that the Sun might have experienced the
dearth of activity in its evolutionary history. And there is no
complete consensus among the solar community whether such grand
minima are chaotic or regular. This work may confirm the
periodicity of the solar secular cycles.

\subsection{51.5-yr cycle ($P_{2}$)}

From the upper panel of Fig.1 we can not get the obvious evidence
of the cycle with period of 51.5 yr. However, the evidence is very
strong in the Fourier power spectra in lower panel of Fig.1. In
fact, from the upper panel of Fig. 1 we may find something of
that, around each peak of the secular cycle, the Schwabe cycles
are in mildly weak amplitudes in ASN. Secular cycles seem to
segment into two sub-peaks. For example, the solar Schwabe cycle A
and 1 around the peak of $G1$, the solar Schwabe cycle 10 around
the peak of $G2$, and the solar Schwabe cycle 20 around the peak
of $G3$. Possibly, these facts are the indicator of the existence
of $P_{2}$ component.

Le \& Wang (2003) investigated the wavelet transformation of ASN
series from 1700 -- 2002, and found the evidence of solar cycles
with period of 11-yr, 53-yr, and 101-yr. In this work, the ASN
series is spanned from 1700 -- 2009, and we rectify the periods as
11-yr, 51.5-year, and 103-yr. Our results are very close to that
of Le \& Wang (2003). Additionally, we find that there is an
interesting phenomenon: $P_{1}/P_{0}=9.36$ is obviously departed
from any integers; however, $P_{1}/P_{2}=2.00$ is fitly equal to
an integer of 2. This evidence shows that $P_{1}$ and $P_{2}$ are
originally connected with each other, and $P_{2}$ seems to be a
second harmonics of $P_{1}$. However, there is no such
relationships between $P_{0}$ and $P_{1}$, $P_{2}$.

Table 1 presents another interesting feature: most of the
asymmetric parameters are less than 1.00 (there are 22 Schwabe
cycles with $Asy<1.00$ among the total 28 cycles, and the
proportion is 78.5\%), the averaged value of asymmetric parameters
is about 0.722, and this implies that most Schwabe cycles are left
asymmetric. They have rapidly rising phases and slowly decay
phases. And the cycle evolution cannot be modelled by some simple
amplitude-modulated sinusoids. Such phenomenon is called as
Waldmeier effect (1961). Among the total 28 cycles, there are only
3 Schwabe cycles with symmetric profiles, and 3 Schwabe cycles
with right asymmetric profiles. The symmetric cycles (No.D, No.5,
and No.16) are very close to the vale of the secular cycles. The
No.7 cycle is a bizarrerie for the super long rise-time (7 years)
and super short decay-time (3 years) with a super large asymmetric
parameter ($Asy=2.33$). Additionally, there is an anti-correlated
relationship between asymmetric parameter and maximum ASN among 28
solar Schwabe cycles. The correlate coefficient is -0.55. i.e. the
stronger the solar Schwabe cycle, the more left asymmetric the
cycle profile.

\section{The Possible Implications of the Solar Cycles}

\subsection{Extrapolation of the Forthcoming Solar Cycles}

It is very important to forecast the forthcoming solar cycles.
Many people make great efforts on this problem (Pesnell, 2008;
Hiremath, 2008; Wang et al, 2009; Strong \& Saba, 2009; Hathaway,
2009; etc). According to the main Features of long-term solar
activity periodicity, we may also make an extrapolation of the
forthcoming solar cycles. Firstly, we note that solar Schwabe
cycle E, 5, and 14 are located in the bottom of secular cycles,
and have rise-time of 4--6 years. At the same time, the solar
Schwabe cycle 24 is also possibly seated in a bottom between $G3$
and $G4$. Then, it is reasonable to suppose that the rise-time of
the solar Schwabe cycle 24 will be in the range of 4 -- 6 years,
i.e., its maximum will occur around 2012 -- 2014. The another
important fact is that all durations of the three bottom solar
Schwabe cycle (cycle E, cycle 5, and cycle 14) are 12 years, which
are relatively long duration Schwabe cycles. Based on the
similarity, we may deduce that the length of Schwabe cycle 24 will
also last for about 12 years, which is longer a bit than the
normal periods of solar Schwabe cycles.

According to variations of the magnitude of solar cycles
(expressed as ASN), we can make some extrapolations for the
forthcoming solar cycles. From Table 1, we know that the averaged
magnitude of ASN among the 28 solar Schwabe cycles during 1700 --
2009 is 105.9 with a standard deviation of 37.8, and the three
bottom cycles (cycle E, 5, and 14) have the ASN magnitude from
47.5 to 63.5. From the above investigations we know that solar
Schwabe cycle 24 may also be a bottom cycle, then we have no other
reason to suppose that solar Schwabe cycle 24 will have an ASN
magnitude of exceeding the range of 47.5 -- 63.5, it is a really
relative weak solar Schwabe cycle. However, this value is lower
than the most results listed in the work of Pesnell (2008), but it
is consistent with the result of Badalyan, et al (2001).

\begin{figure}
\begin{center}
 \includegraphics[width=8.0cm]{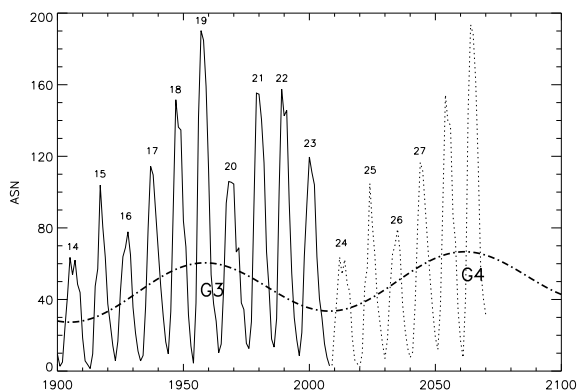}
  \caption{The extrapolations of the forthcoming secular cycle 4 (G4), the solar Schwabe cycle 24 and the beyond. }
\end{center}
\end{figure}

Fig. 2 plotted the extrapolation of the forthcoming solar Schwabe
cycles by the similarity of the trend-line induced from Fig.1. It
shows that the position of solar Schwabe cycle 24 is very close to
the valley between the secular cycle $G3$ and $G4$, which is very
similar to the solar Schwabe cycle 5 in the valley between $G1$
and $G2$, and the solar Schwabe cycle 14 in the valley between
$G2$ and $G3$. So it is very reasonable to extrapolate that the
solar Schwabe cycle 24 will be similar to that of the solar
Schwabe cycle 5 and cycle 14. This similarity implies that the
Schwabe cycle 24 will be a relatively weak activities.

Based on these similarities and Equ.(1), we may plot the
extrapolated ASN of the subsequent solar Schwabe cycles and the
secular cycle $G4$ in Fig.2. From these extrapolations, we may
find that solar Schwabe cycle 24 will reach to the apex in the
year of about 2012 - 2014, and this cycle may last for about 12
years or so. It will be a long, but relatively weak Schwabe cycle.

\subsection{Distributions of Solar Flare Events in a Schwabe Cycle}

When and what kind of solar flare events will occur is also an
intriguing problem. Fig.3 presents comparisons between ASN and the
distributions of the appearance rate of solar flares and the
annual averaged solar radio flux (arbitrary unit) at frequency of
2.84 GHz around the solar Schwabe cycle 23. The appearance rate of
C-class, M-class, and X-class (limit from X1.0 to X9.9) flares are
presented by their annul flare numbers, and the appearance rate of
the super-flares (here we define the super-flare as $\geq$ X10.0)
are presented by their related GOES soft X-ray class. The vertical
dashed line marks the time of magnetic maximum of solar Schwabe
cycle 23. The 2.84 GHz solar radio flux is observed at Chinese
Solar Broadband Radiospectrometer (SBRS/Huairou) (Fu et al, 2004).
The data set of the annul flare numbers is compiled from GOES
satellite at soft X-rays (http://www.lmsal.com/SXT/homepage.html).

\begin{figure}
\begin{center}
 \includegraphics[width=8.2cm, height=8.0cm]{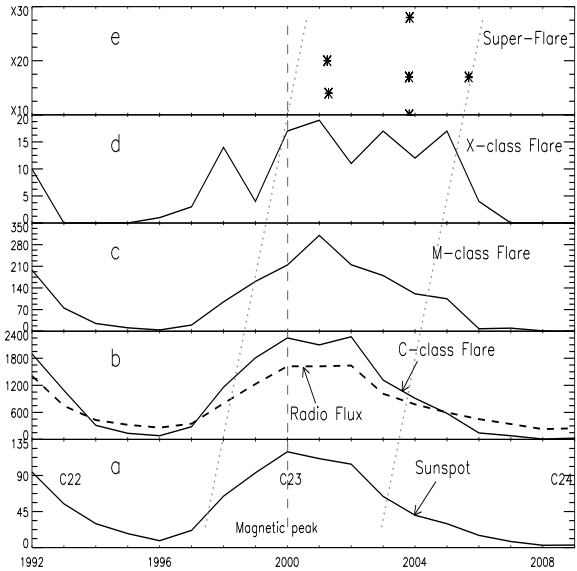}
  \caption{Comparisons between ASN (a) and the distribution of the strength of solar flares
  and the annual averaged solar radio flux (the dashed curve in panel (b), in arbitrary unit) at frequency of 2.84 GHz around the solar Schwabe cycle 23.
  The appearance rate of C-class (b), M-class (c), and X-class flares (d) are presented by the annul flare numbers, and the appearance rate of the super-flares
  are presented by the related GOES soft X-ray class (e). The vertical dashed line marks the time of magnetic maximum of cycle 23.
  The two tilted-dotted lines show the trend of the flare distribution.}
\end{center}
\end{figure}

From Fig.3 we may find an obvious tendency: the more powerful
solar flares are inclined to occur in the period after the maximum
of solar Schwabe cycle (presented by the two tilted-dotted lines).
For example, the ASN of solar Schwabe cycle 23 reaches to its
maximum in about 2000, and the annual number of C-class flares and
M-class flares reach to the maximum in about 2001, while the
annual number of X-class flare reaches to its apex in about 2002,
the most concentration of the super-flares is occurred in 2003.
During the whole Schwabe cycle 23 (1996 -- 2008) there are 12995
C-class flares, 1444 M-class flare, 119 X-class flares, and 6
super-flares. Among them there are 7414 C-class flares (57.1\%),
950 M-class flare (65.8\%), 80 X-class flares (67.2\%) and all the
super-flares are occurred after the year of ASN apex (2000) of the
cycle. In 2005, which is very close to the magnetic minimum of
cycle 23, there are 18 X-class flares occurred. Additionally, the
annual averaged solar radio flux (in panel b) at frequency of 2.84
GHz are very similar to the profile of the annual C-class flare
numbers. Generally, the solar radio emission at 2.84 GHz is mainly
associated with the non-thermal eruptive processes. It shows that
all of the 6 super-flares ($>X10$) occurred in the decay phase of
the Schwabe cycle 23, including the the largest flare event (X28,
2003-11-03) recorded in NOAA so far. These facts imply that the
stronger flare events are inclined to occur in the decay phase of
the Schwabe cycle. However, so far we can not make a doubtless
conclusion because we have no enough reliable observation data of
the other Schwabe cycles.

The above characteristics of the solar flare distribution in solar
Schwabe cycles is also implied the asymmetric properties of the
Schwabe cycles. However, it is most intricate that most of the
powerful flares are occurred not in the rising-phase but in the
decay-phase.

\subsection{Discussions on Solar Cycle Mechanism}

In this work, we make an assumption that future behaviors of the
solar activity in upcoming several decades years can be deduced
from the averaged behaviors in the past several hundred years. And
this can be accepted because the several hundred-year is so much
short related to the life-time of the Sun when it is at the main
sequence. We may regard the Sun as a steady-going system which
will run according to its averaged behaviors of its past several
hundred years and last for several hundred years again.

The time-scales of the solar 11-yr Schwabe cycle and 103-yr
secular cycles are much shorter than the diffusion time-scale of
solar large-scale global magnetic field structures (about
$3\times10^{7}$ yr, Stix, 2003), and much larger than the solar
dynamical time-scale (for example, the 5-min oscillations, etc).
The origin of these cycles is a remarkable unsolved tantalizing
problem. We need to explore some theoretical models which may
carry the energy from the interior to the surface and release in
solar atmosphere in some periodic forms. Presently, there are two
kinds of theoretical models for solar cycles, one is turbulent
dynamo models, the other is MHD oscillatory models. However, both
of them have much difficulties to interpret the main features of
solar cycles reliably (Hiremath, 2009).

As for the 11-yr Schwabe cycle, in spite of much of difficulties,
possibly dynamo theory is the most popular hypothesis to explain
the generations (Babcock, 1961; Leighton, 1964, 1969). This model
finds its origin in the tachocline where strong shearing motions
occur in the solar internal plasma, store and deform the magnetic
field, and can give a semi-empirical model of solar 11-yr cycle
and reproduced the well known sunspot butterfly diagrams. At the
same time, there are several other approaches. Polygiannakis \&
Moussas (1996) assume that the sunspot-cycle-modulated irradiance
component may be caused by a convective plasma current, this
current can drive a nonlinear RLC oscillator. This model can
describe the shape and the related morphological properties of the
solar cycles, and give some reasonable interpretations to the long
period inactive Maunder Minimum.

The closeness between the 11-yr Schwabe cycle and the 11.86-yr
period of Jupiter has been noted for a long time. And this let us
to speculate that planetary synodic period may resonate with solar
activity (Wood, 1975). Grandpierre (1996) proposed that the
planetary tides of co-alignments can drive the dynamo mechanism in
the solar interior and trigger solar eruptions. They calculate the
co-alignments (conjunction and opposition) of the Earth, Venus,
and Jupiter and find that their co-alignment periods are in the
range of 8.7588 -- 13.625 yr, the averaged value is around 11.2
yr, being very close to the observed Schwabe cycle of 11-yr.
However, de Jager \& Versteegh (2005) compare the accelerations
due to planetary tidal force, the Sun's motion around the
gravitational center of the solar system, and the observed
acceleration at the level of tachocline, and find that the latter
are by a factor of about 1000 larger than the former two, and
assert that the planetary tidal force can not trigger the dynamo
mechanism observably. However, in my opinion, as the plasma system
is always very brittle, even with a very small perturbation, a
variety of instabilities are also very easy to develop,
accumulate, and trigger a variation as the solar Schabe cycle.
Hence, at present stage, it is very difficult to confirm which
factor is the real driver of solar Schwabe cycles.

The most bewildering is the harmonic relationship between the
long-term cycles $P_{1}$ and $P_{2}$. This may imply that solar
long-term active cycles have wave's behaviors. However, so far, we
don't know more natures of this kind of cycles. The dynamo theory
can give some reasonable interpretation to the 11-yr solar Schwabe
cycle, but it can not present even if a plausible explanation to
the 51.5-yr cycle and 103-yr secular cycle so far. In 1943, Alfven
assumed that solar large-scale dipole magnetic field has the axis
coincide with the rotation axis, and the magnetic disturbance in
the interior travels along the field line to the surface at Alfven
speed. This magnetic disturbance will excite MHD oscillation and
transport to the whole Sun, the travel time of the MHD oscillation
is about 70 -- 80 yr. This time scale is neither agree with the
11-yr period, nor with the time scale of the secular cycles.

Many people believe that the 11-yr solar Schwabe cycle is
originated from the tachocline which is a thin shear layer of
strong differential rotation motion at the base of the solar
convection zone (Dikpati, 2006). Helioseismology suggests that the
tachocline in thin of the order of 1\% of the solar radius
(Christensen-Dalsgaard \& Thompson, 2007). It is well-known that
the Kelvin-Helmholtz instability is very easy to develop around a
layer with strong shear motion. From the work of Gilliland (1985),
we know that the time scale of Kelvin-Helmholtz instability for
the whole Sun is $3\times10^{7}$ yr. Then it is possible to assume
that the time scale of Kelvin-Helmholtz instability around the
thin tachocline can be reduced to the order of 100-yr. Rashid,
Jones, \& Tobias (2008) pointed out that the evolution of the
hydrodynamic instability of the slow tachocline region will occur
on timescale of hundred years. However, we need much of
investigations to confirm this assumption and to understand the
long-term solar cycles. The origin of long-term solar cycles is
still a big unsolved problem.

\section{Conclusions}

From the above analysis and estimations, we obtain the following
conclusions:

(1) besides the well-known 11-yr solar Schwabe cycle, two
long-term solar cycles, 103-yr secular cycle (Gleissberg cycle)
and the 51.5-yr semi-secular cycle are identified;

(2) the solar Schwabe cycle 24 is in a vale between the two
secular cycles (G3, and G4), it will be a relatively weak and long
active cycle, which may reach to its apex in about 2012-2014 and
last for about 12 years;

(3) most solar Schwabe cycles are left asymmetric, they have
rapidly rising phases and slowly decay phases. The most intriguing
is that most of the solar powerful flares occurred in the
decay-phase of the solar Schwabe cycle.

(4) the secular cycle is possibly associated with the solar inner
large scale motions as well as the dynamo processes can be account
for the 11-yr Schwabe cycles. However, there are much of works
need to do to understand the mechanism of the secular cycles.

\acknowledgments

The author would like to thank the referee's valuable comments on
this paper. This work was supported by NSFC Grant No. 10733020,
10873021, CAS-NSFC Key Project (Grant No. 10778605), and the
National Basic Research Program of the MOST (Grant No.
2006CB806301).

\label{}

\end{document}